%% file: main.tex
\documentclass[letterpaper,twocolumn,10pt]{article}
\usepackage{usenix,epsfig}

\date{}

\usepackage[babel]{csquotes}
\usepackage{enumerate}
\usepackage{authblk}
\usepackage{tabularx}
\usepackage{multirow}
\usepackage{ctable}
\usepackage{listings}
\usepackage{graphicx}
\usepackage{subfig}
\usepackage{subfloat}
\usepackage{pifont}
\usepackage{comment}
\usepackage{multirow}
\usepackage{float}
\usepackage[bottom]{footmisc}
\usepackage[utf8x]{inputenc}
\DeclareCaptionType{copyrightbox}
\restylefloat{table}

\title{Isolario: a Do-ut-des Approach to Improve the Appeal of BGP Route Collecting}

\author{Enrico Gregori}
\author{Alessandro Improta}
\author{Luca Sani}
\affil{Institute of Informatics and Telematics, Italian National Research Council\\Pisa, Italy}

\newcolumntype{x}[1]{%
>{\centering\hspace{0pt}}p{#1}}%

\newcommand{\mytilde}{\raise.17ex\hbox{$\scriptstyle\mathtt{\sim}$}}

\begin{document}	
	\maketitle
	\input{abstract}
	\input{introduction}
	\input{related_work}
	\input{architecture}
	\input{need_for_speed}

	\input{apps}

	\input{conclusions}

	{\footnotesize \bibliographystyle{acm}
        \bibliography{references}}
\end{document}

%% file: abstract.tex
\subsection*{Abstract}
The incompleteness of data collected from BGP route collecting projects is a 
well-known issue which potentially affects every research activity carried out on
the analysis of the Internet inter-domain routing. Recent works explained that 
one of the possible solutions is to increase the number of ASes feeding these
projects from the Internet periphery, in order to reveal the hidden portion of peering
connectivity of their upstream providers. The main problem is that these 
projects are currently not appealing enough for the network administrators of 
these ASes, which are typically not aware of their existence or not interested 
enough to share their data. Our contribution is Isolario, a project based on the
do-ut-des principle which aims at persuading network administrators to share 
their routing information by offering services in return, ranging from real-time
analyses of the incoming BGP session(s) to historic analyses of routing 
reachability. To the best of our knowledge, Isolario is the only route 
collecting project publicly available which offers a set of services to its 
users to encourage their participation, aiming at increasing the amount of BGP 
data publicly available for research purposes.

%% file: introduction.tex
\section{Introduction}
Route collectors deployed by the Route Views project of the University of 
Oregon~\cite{RouteViews} and the Routing Information Service (RIS) project of
the Réseaux IP Européens Network Coordination Center (RIPE NCC)~\cite{RIS} have
been an invaluable source of information for researchers all over the world for 
the past twenty years. Data collected from these projects greatly contributed to
shed light on the knowledge concerning the Internet structure (e.g.~\cite{Chen02,
Gao01, Li04, Moore03, Yegneswaran03}, to mention just a few), which has been 
lost the day after the NSF's privatization in 1995, when regional 
networks started to buy national-scale Internet connectivity from private 
long-haul networks growing since the early 90s~\cite{Leiner09, Norton14}. 
Route collectors are nothing more than simple servers that mimic the role of border
routers and which establish sessions with various organisations using the Border
Gateway Protocol 4 (BGP)~\cite{RFC4271}. BGP is the de-facto standard protocol 
used in inter-domain routing, and it is used to establish logical 
connections between pair of Autonomous Systems\footnote{An Autonomous System is 
defined as "\textit{a set of routers under a single technical administration}", 
which administration "\textit{appears to other ASes to have a single coherent 
interior routing plan and presents a consistent picture of what networks are 
reachable through it"~\cite{RFC1930}}} (ASes) to exchange reachability 
information to be used in the BGP decision process of the ASes involved. 
Those pieces of information are carried in \texttt{UPDATE} messages in the form of
path attributes, each with its own meaning in the routing process. Route 
collectors behave like one of the two ends of the logical connection, but they 
only collect incoming BGP messages without generating any routing traffic 
directed to the other party. Thus, they are able to receive in real-time the 
best routes chosen by the BGP decision process of the connected AS Border Router (ASBR),
and data collected allows to study the routing characteristics of 
the connected AS. Needless to say, these attributes represent a potential gold 
mine of information about the Internet ecosystem for researchers, and they can 
be found in Multi-threaded Routing Toolkit (MRT) export format~\cite{RFC6396} 
files in the website of each project. \\

However it is not all a bed of roses. Data collected by Route Views and RIS is 
indeed known to be largely incomplete~\cite{Gregori12, Oliveira10}. Moreover, 
most of the organisations which decided to participate in these projects are 
very large Internet Service Providers (ISPs)~\cite{Gregori12} and, as a 
consequence, data collected by these projects could lead to biased results, 
depending on the analysis carried out. For example, they fail to reveal 
the largest part of the peering ecosystem~\cite{Chatzis13, He09} mostly due to 
the nature of their participants, which lead several researchers to believe that
the Internet was following a power-law node degree distribution when analyzed at
the AS-level abstraction~\cite{Albert02, Barabasi99, Faloutsos99}. We believe 
that the main cause of these drawbacks is the voluntary basis on which these 
projects rely on. In our opinion most of the largest ISPs are attracted by the 
opportunity to exhibit their interconnections to potential customers, while the 
administrators of the smallest organisations may not find any strong motivation
to join a route collecting project. \\

To overcome this polarization and the lack of routing information we developed 
Isolario, a non profit research project which aims to improve the knowledge 
about the Internet ecosystem by enhancing the appeal of the classic concept of 
BGP route collector. To do that, Isolario provides a set of services to the network 
administrators of each AS participating aimed at easing their jobs. These services 
are built on top of the BGP flow which is collected, analysed and finally shared 
with the research community. To the best of our knowledge, Isolario is the only research 
project publicly available which tries to push network administrators to participate 
and increase the number of data sources of BGP data available for research purposes 
in addition to mere route collecting. In detail, this paper 
describes the architecture of Isolario, focusing in particular on the design 
choices and the challenges engaged to build a route collecting system able to 
support the implementation of real-time and historical data services. Finally, we describe briefly
the set of services built on top of either the route collecting software or the 
archival system, some of which are already available on 
Isolario. With these services a network administrator 
participating to Isolario would be able, for example, to detect in 
\textit{real-time} pathological inter-domain routing events like route flapping
without increasing the computational load on its routers or introducing 
third-party software. \\

The rest of the paper is organised as follows. Section~\ref{sec:related_work} 
describes the state of the art on BGP collecting systems publicly available. 
Section~\ref{sec:architecture} introduces the Isolario system and its 
architecture while Section~\ref{sec:need_for_speed} describes the route 
collecting engine and the archival system. Section~\ref{sec:apps} focuses on the
services developed so far, introducing examples of both real-time and historic
analysis services. Finally, Section~\ref{sec:conclusions} concludes the paper.

%% file: related_work.tex
\section{Related work}
\label{sec:related_work}

The art of BGP route collecting dates back to 1997, when data collection was
performed by the Measurement and Operations Analysis Team (MOAT) of the National
Laboratory for Advanced Network Research (NLANR) project~\cite{McGregor00}. 
Since then, the concept of route collecting has not changed significantly. Route 
collection is performed by servers which mimic the behaviour of an ASBR using a 
routing suite like Quagga~\cite{Quagga}. Route collectors 
establish BGP sessions with organisations which voluntarily agree to 
participate, and regularly dump routing information. Data was initially 
extracted via shell scripting, dumping regular Routing Information Base 
(RIB) snapshots. Then, thanks to the introduction of MRT~\cite{RFC6396} the 
data format has been standardized and every single packet in BGP sessions 
established towards route collectors was caught and stored, thus making available
the possibility to re-create the BGP flow collected in a given period of time to 
investigate network issues. NLANR has ceased its activities some years 
ago\footnote{http://www.nlanr.net/}, but route collecting was carried on by the
Route Views project at University of Oregon~\cite{RouteViews} -- which collects MRT 
data since 2001 -- and by the Routing Information Service (RIS) at the Réseaux IP 
Européens Network Coordination Center (RIPE NCC)~\cite{RIS} -- which collects MRT
data since 1999. Both projects made publicly available periodic RIB snapshots and a 
collection of every single BGP packet collected at different time 
intervals\footnote{Route Views provides RIB snapshots every 2 hours, RIS 
every 8 hours. Route Views dumps \texttt{UPDATE} messages every 
15 minutes, RIS every 5 minutes.}. \\

\begin{figure*}[!t]
    \includegraphics[width=\textwidth,natwidth=2210,natheight=986]{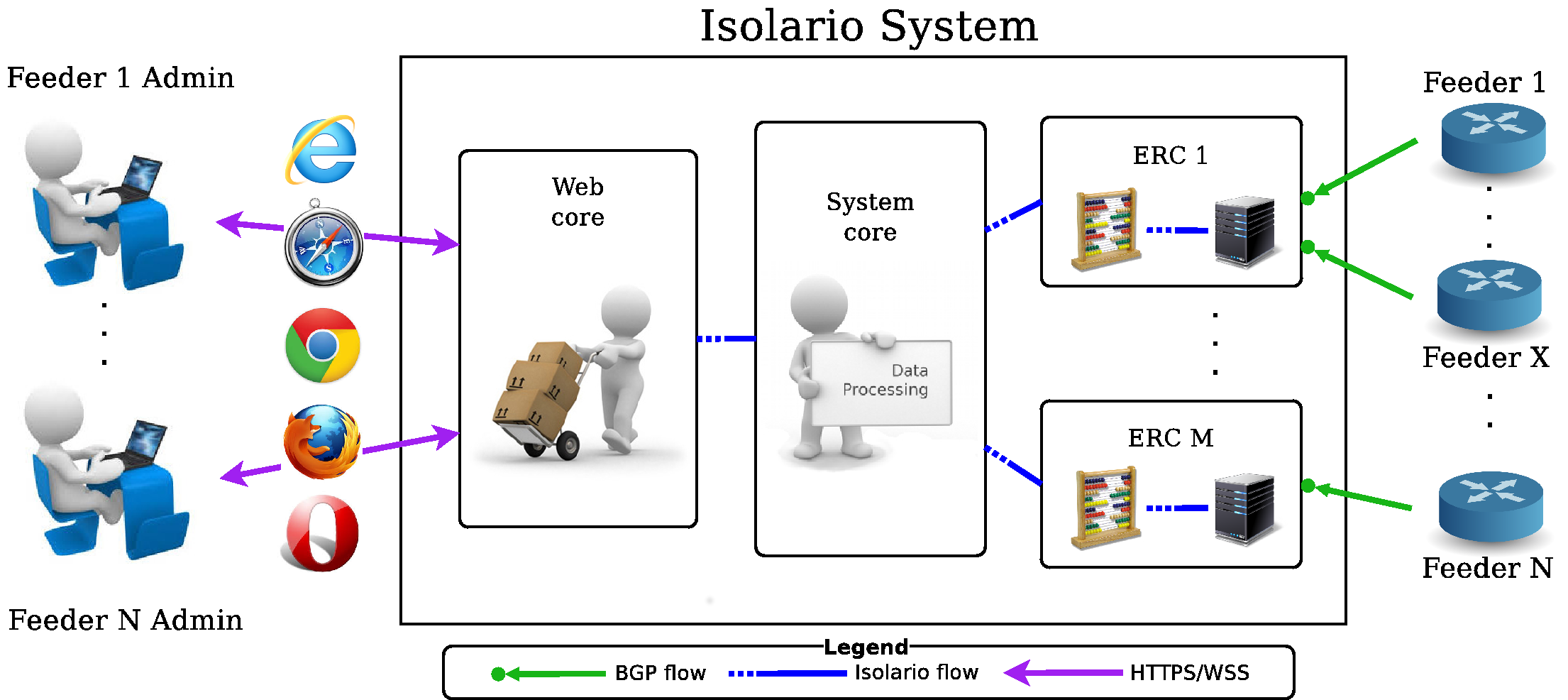}
    \caption{Isolario system overview}
    \label{fig:isolario_architecture}    
\end{figure*}

A step forward from the classic concept of BGP route collector has been achieved with
BGPmon~\cite{BGPmon}, a tool developed at Colorado State university to 
enable the real-time monitoring of BGP routing information in addition to simple
route collecting. The idea is to replace classic routing suites with
dedicated software which is able to provide a public live BGP data stream
in XML format, in addition to being able to mimic a BGP border router and 
store routing information. This flow can be exploited to perform a wide range of experiments 
and analyses, as well as homemade BGP monitoring systems. However, even though 
this idea is extremely interesting, up to date not many organisations have decided to 
join this project\footnote{http://bgpmon.netsec.colostate.edu/peer-list.html}, 
and only a few of those who joined announce a full routing 
table~\cite{Gregori12}. In our opinion, the main reason of this phenomenon is 
that BGPmon still fails to provide a strong motivation to potential data sources 
to feed their project with their BGP routing information.
By providing a real-time flow, BGPmon increases only the 
appeal of data and the amount of potential data users, while the appeal for 
potential BGP data sources remains unaltered. \\

We believe that a good idea to 
attract new BGP data sources would be to provide a set of services to
each network administrator who decides to participate which could be
useful to manage their network in \textit{real-time} and/or to search their
\textit{historic} routing data to investigate for network issues. To the best of our
knowledge, up to date no route collecting project pursed a similar approach. RIPE NCC has done 
something in this direction with their project RIPEstat~\cite{RIPEstat}, but the 
data is freely provided to public, without any need to connect an ASBR to their RIS 
project. RIPEstat is indeed a toolbox that ease the access to the various datasets maintained at RIPE NCC -- 
such as DNS and RIS routing data -- and offers a list of widgets to analyse any 
public network composing the Internet, with additional information for all those 
networks registered in RIPE database. A similar tool has been developed by the 
Internet Research Lab of UCLA with the Cyclops project~\cite{Chi08}, which do 
not collect any BGP route but uses external data source (e.g. looking glasses, 
route servers and BGP data from Route Views and RIS) to compare the behaviour 
of a given network as observed by data collected with the behaviour intended by 
the AS administrator. Both tools provide up to date only \textit{a posteriori} 
services, since they both exploit routing data dumps and can only be used in 
\textit{a posteriori} analyses, i.e. after a problem has occurred.

%% file: architecture.tex
\section{Architecture overview}
\label{sec:architecture}

Isolario is a distributed system devised to collect, parse, elaborate 
BGP data sent from the ASBRs of its participants  -- hereafter 
\textit{feeders} -- and to provide results to network administrators of the 
participating ASes -- hereafter \textit{Isolario users} -- by introducing the 
minimum amount of delay as possible. To achieve that, Isolario makes use 
of three main components, as depicted in Figure~\ref{fig:isolario_architecture}: 
\textit{i}) Web core, \textit{ii}) System core, and \textit{iii}) Enhanced Route
Collectors (ERCs). Each component is designed to be modular and scalable, to
allow the introduction of new pieces of equipment in a plug-and-play fashion 
without affecting the user experience. In addition to that, components are 
interconnected via TCP, in order to allow the deployment of part of the system 
in different parts of the world. Isolario system has been designed to fulfill
two main purposes. On one side, it must be able to establish and maintain BGP sessions with 
each feeder, collecting at the same time routing information in the MRT repository. On 
the other side, it must be able to react to user service input requests by computing and 
elaborating BGP incoming packets or historic data and providing the requested 
service output(s). \\

\begin{figure}[!b]
    \includegraphics[width=\columnwidth,natwidth=2174,natheight=1019]{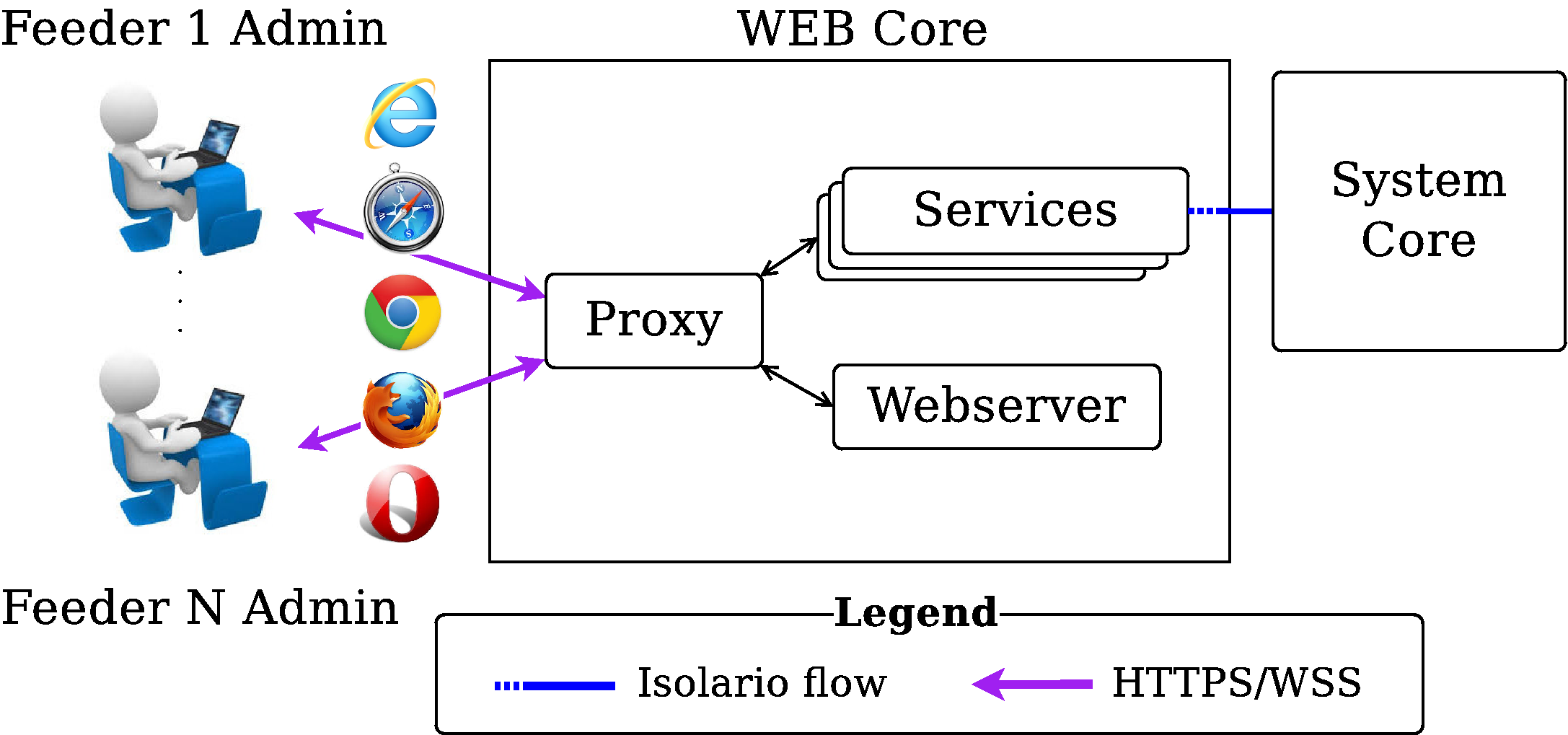}
    \caption{Web core architecture overview}
    \label{fig:web_core}    
\end{figure}

An Isolario user interacts with the \textit{web core} component 
(Figure~\ref{fig:web_core}) from a private area of the Isolario website, where 
every service available can be accessed. As soon as a service has been chosen, it 
is created a dedicated WebSocket~\cite{RFC6455} session between the Web core and the user web 
browser on which user requests and results will flow in real-time. WebSockets 
are chosen to allow the server to send data directly to the user browser as soon
as new data is available, avoiding costly and unuseful polling traffic being
generated by the web browser or explicit refresh requests from the user. On
the Web core side, HTTPS and WebSocket flows are dispatched by a proxy module
respectively to the Webserver or to the related service module which will 
propagate the traffic to the system core. To minimize the amount of 
unuseful routing data in the system, each service also identifies which 
portion of IPv4/IPv6 space is interested by the user query, and generates 
dedicated messages to configure filters on the proper ERC(s). \\

\begin{figure}[!b]
    \includegraphics[width=\columnwidth,natwidth=2046,natheight=856]{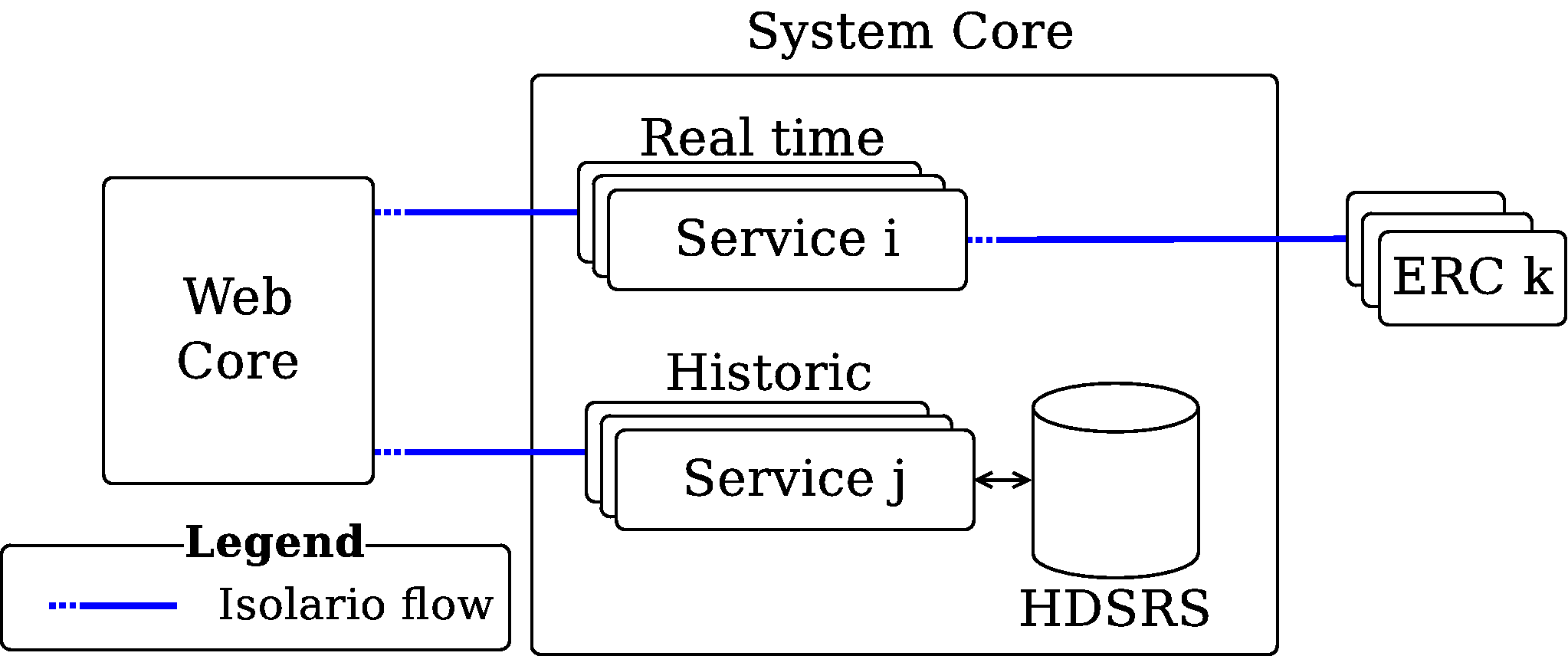}
    \caption{System core architecture overview}
    \label{fig:system_core}    
\end{figure}

The \textit{system core} component (Figure~\ref{fig:system_core}) then handles 
user requests propagated by the web core as well as filter configuration 
messages. This component is composed by a set of real-time service modules
and a set of historic service modules. The real-time service modules 
dispatch user requests and filter configuration messages to the related 
service modules located on ERCs, and aggregates results based on the partial
results received from each ERC (if required). On the opposite, historic
services fetch data from the Historic Data Storage and Retrieval System
(HDSRS), which is designed to allow fast access to stored routing data. 
Details about the HDSRS can be found in Section~\ref{sec:hdsrs}.\\

\begin{figure}[!t]
    \includegraphics[width=\columnwidth,natwidth=2883,natheight=1522]{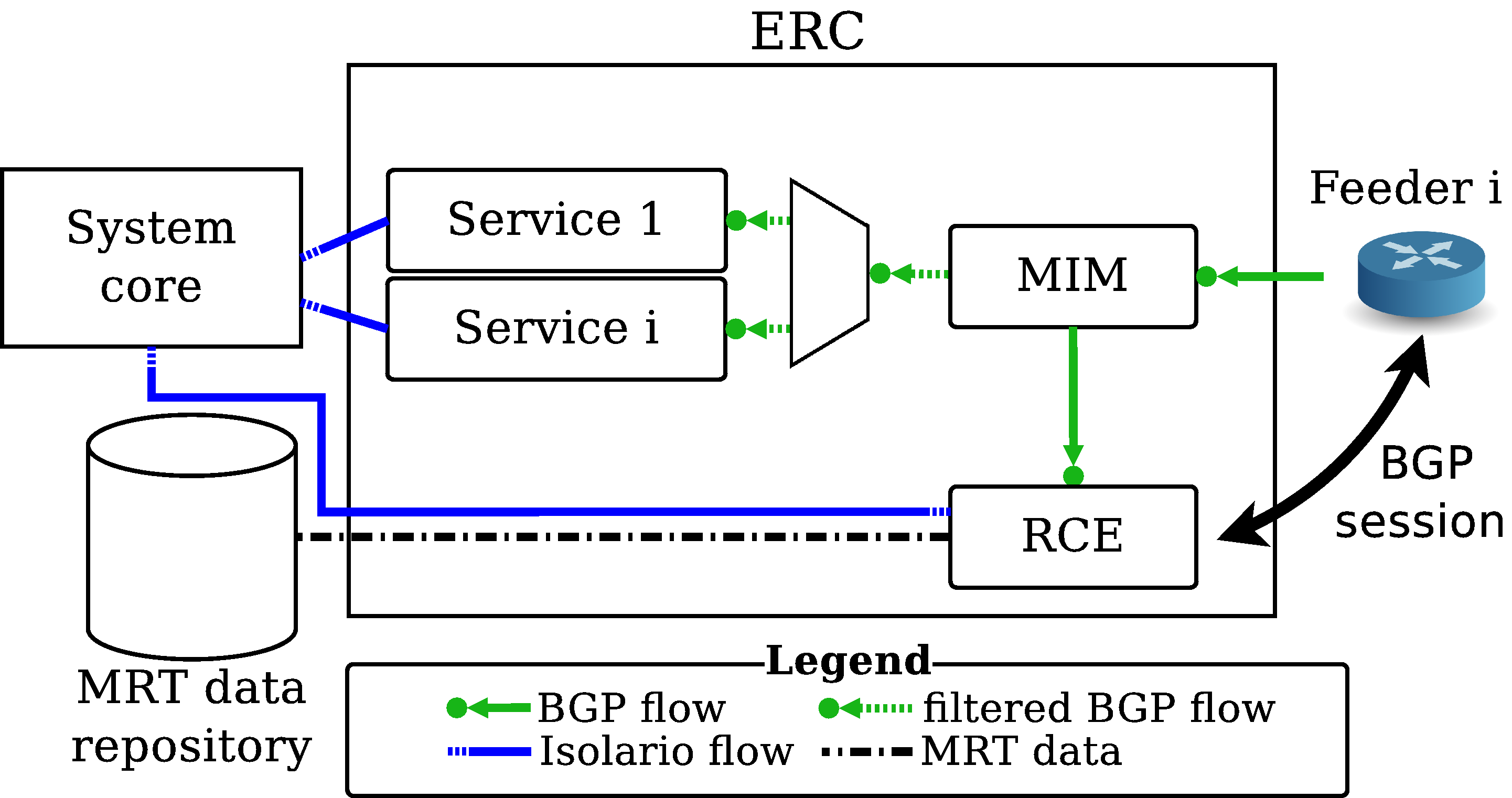}
    \caption{Enhanced route collector overview}
    \label{fig:route_collector}    
\end{figure}

Finally, each \textit{ERC} has the role to establish 
and maintain active BGP sessions towards a set of feeders, dispatching every 
single BGP packet which matches filtering criteria towards service modules. 
Every BGP session on ERCs is maintained by the \textit{Route Collecting Engine 
(RCE)}, which also shoots every 2 hours a snapshot of its RIB and dumps every 5 
minutes the full sequence of BGP packets received in the last 5 minutes in MRT 
format. RCE details are provided in Section~\ref{sec:rce}. To inject the incoming
BGP packets into the system as soon as possible, we introduced a dedicated 
\textit{Man In the Middle (MIM)} module between each feeder and our system. This 
module forwards the BGP packets to the RCE and, at the same time, applies the filter configuration 
messages received from the system core. Every single incoming BGP packet 
matching the filter criteria is then forwarded also to the proper service 
module, which will perform all the required elaborations. Results obtained from 
service modules  are finally propagated back to the user through the 
Isolario system and the WebSocket session.

%% file: need_for_speed.tex
\section{Towards acceptable user experience}
\label{sec:need_for_speed}

Isolario system is founded on the reciprocal usefulness criterion, also known as 
the \textit{do-ut-des} principle. This principle is known since 
Roman ages and has been used as political motto by several statesmen in history, 
such as Constantine the Great. To apply this principle successfully, the ruler typically 
\textit{gave} something valuable like booty or land, \textit{in return for} 
loyalty and armed support by the retinue. In Isolario's case, this principle
says that to attract as many AS administrators as possible, Isolario has to 
provide as many useful and interesting (and thus \textit{valuable}) services as 
possible, still guaranteeing to every Isolario user the best user experience. In 
this sense, Isolario system has been designed with particular focus on reducing
data processing delays. In this section we describe two of the most important 
parts of the system, which allow users to receive results as soon as possible:
the \textit{Route Collecting Engine} and the \textit{Historic Data Storage
and Retrieval System}.

\input{real-time}

\input{non-real-time}

%% file: real-time.tex
\subsection{Route Collecting Engine}
\label{sec:rce}

\begin{figure}[!t]
    \includegraphics[width=\columnwidth,natwidth=1383,natheight=1024]{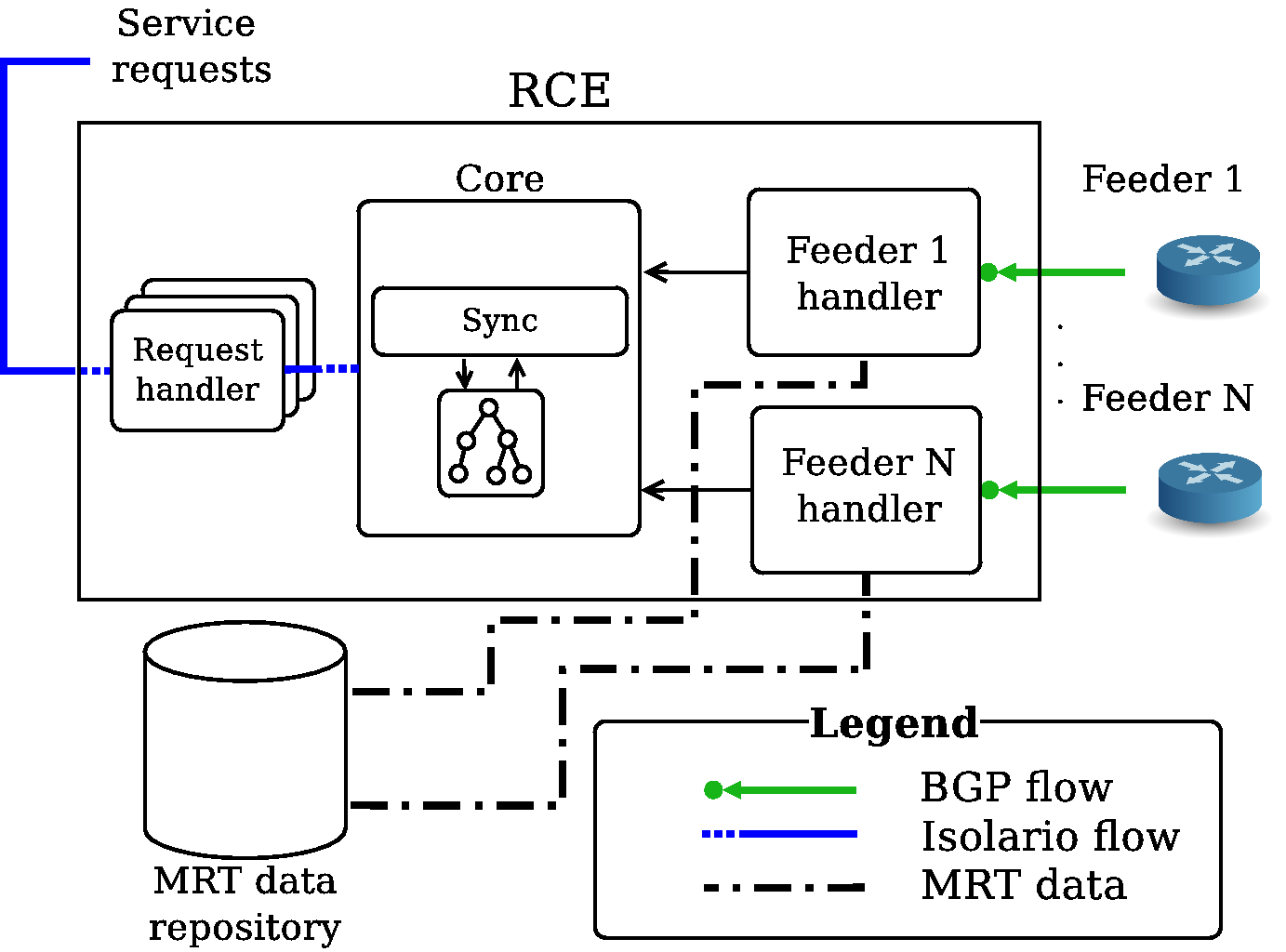}
    \caption{Route Collecting Engine (RCE)}
    \label{fig:rce}
\end{figure}

One of the main characteristic of Isolario is the availability of
services dedicated to the real-time analysis of the incoming BGP flows, either 
focused on portions of the Internet routing table or focused on the reachability
of a given network prefix from every perspective available in Isolario (see 
Section~\ref{sec:apps}). Services belonging to this set share two common main 
requirements: \textit{a}) BGP data 
has to be received by interested services as soon as they arrive to any route 
collector, without any additional delay, and \textit{b}) services 
have to be able to access the RIB of route collectors as soon as possible to 
retrieve the BGP path attributes required at any time. While the first requirement is satisfied 
by design thanks to the MIM module described in Section~\ref{sec:architecture}, 
the second requires a dedicated route collecting engine able to maintain over 
time the best routes received from each feeder, store them in MRT files and, at 
the same time, able to provide the status of any of the stored Network Layer 
Reachability Information (NLRI) with the minimum delay as possible to any 
requesting service. To the best of our knowledge, the requirements 
described above cannot be completely fulfilled by any of the routing suites 
publicly available. On one side, general purpose routing suites (e.g. 
Quagga~\cite{Quagga}, Bird~\cite{Bird}, XORP~\cite{XORP} and exabgp~\cite{exabgp})
implement routing functionalities that are not needed for the purpose of 
route collecting (e.g. BGP decision process and proper import/export policies), 
introducing unnecessary computation overhead and delays. On the other side, routing
suites efficient in terms of RAM and CPU load which would
allow fast access to RIB content from external services do not have built-in 
support for generating periodic dumps of RIB and updates in MRT format~\cite{RIS_research_RC}. 
To completely satisfy the above requirements we created a dedicated component 
named \textit{Route Collecting Engine} (see Figure~\ref{fig:rce}). This component is \textit{a}) scalable 
with respect to the number of feeders, \textit{b}) has full support of MRT 
dumps (periodic RIB snapshots and collections of BGP \texttt{UPDATE} messages 
received over time), and \textit{c}) allows services to safely read the 
content of the RIB of route collectors while its content is modified by incoming 
BGP packets.

%% file: non-real-time.tex
\subsection{Historic Data Storage and Retrieval System}
\label{sec:hdsrs}

\begin{figure}[!t]
    \includegraphics[width=\columnwidth]{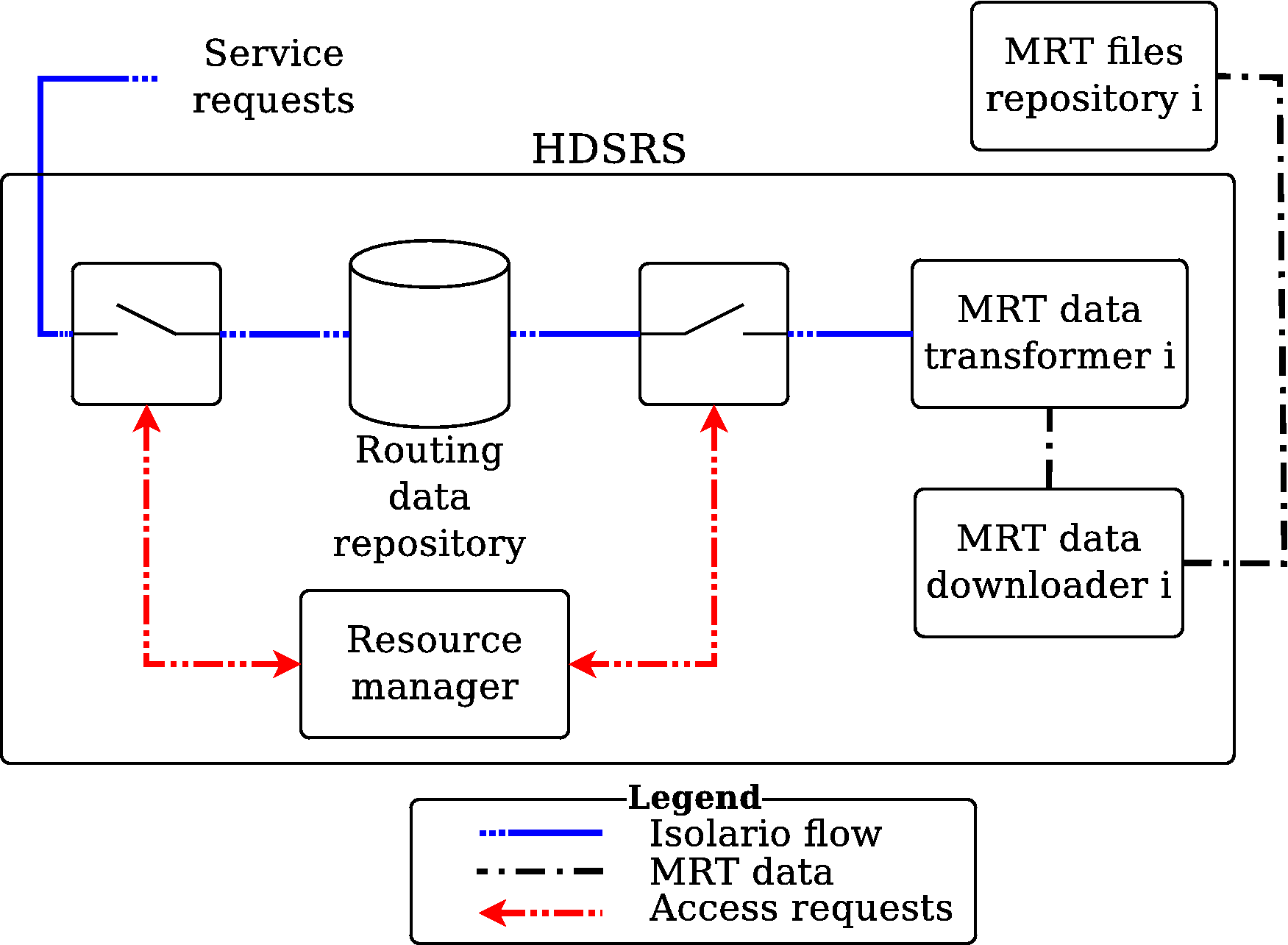}
    \caption{Historic Data Storage and Retrieval System (HDSRS)}
    \label{fig:hdsrs}
\end{figure}

Differently from real-time services, the only requirement of historic services 
is that the system has to provide fast access to historic data in order to 
retrieve results to the users' web browser as soon as possible. In particular, 
data access speed should be neither route-dependent nor time-dependent, i.e. 
routing data should be accessible with complexity $\mathcal{O}(1)$ from the 
dedicated handler \textit{independently} from the amount of information stored 
in the system. Routing data is typically maintained in MRT files (e.g. Route 
Views, RIS). MRT files are time sorted sequences of BGP packets collected by 
multiple sources and which contain information related to the (potentially) 
whole IPv4/IPv6 space as announced by each BGP peer monitored. Even though it is
possible to organize file names to ease the access 
to obtain routing data related to the time span requested by the service, it
is still required to read the whole file to discern information related to 
the route(s) under analysis from the rest of the data. More importantly, this 
has to be done every time an historic service requests routing information, 
making this approach to be unfeasible to our ends. To guarantee fast replies to
service requests, we designed the \textit{Historic Data Storage and Retrieval 
System (HDSRS)} (see Figure~\ref{fig:hdsrs}), which is based on a new logical
file organisation. 

%% file: apps.tex
\begin{figure*}[!t]
    \includegraphics[width=\textwidth]{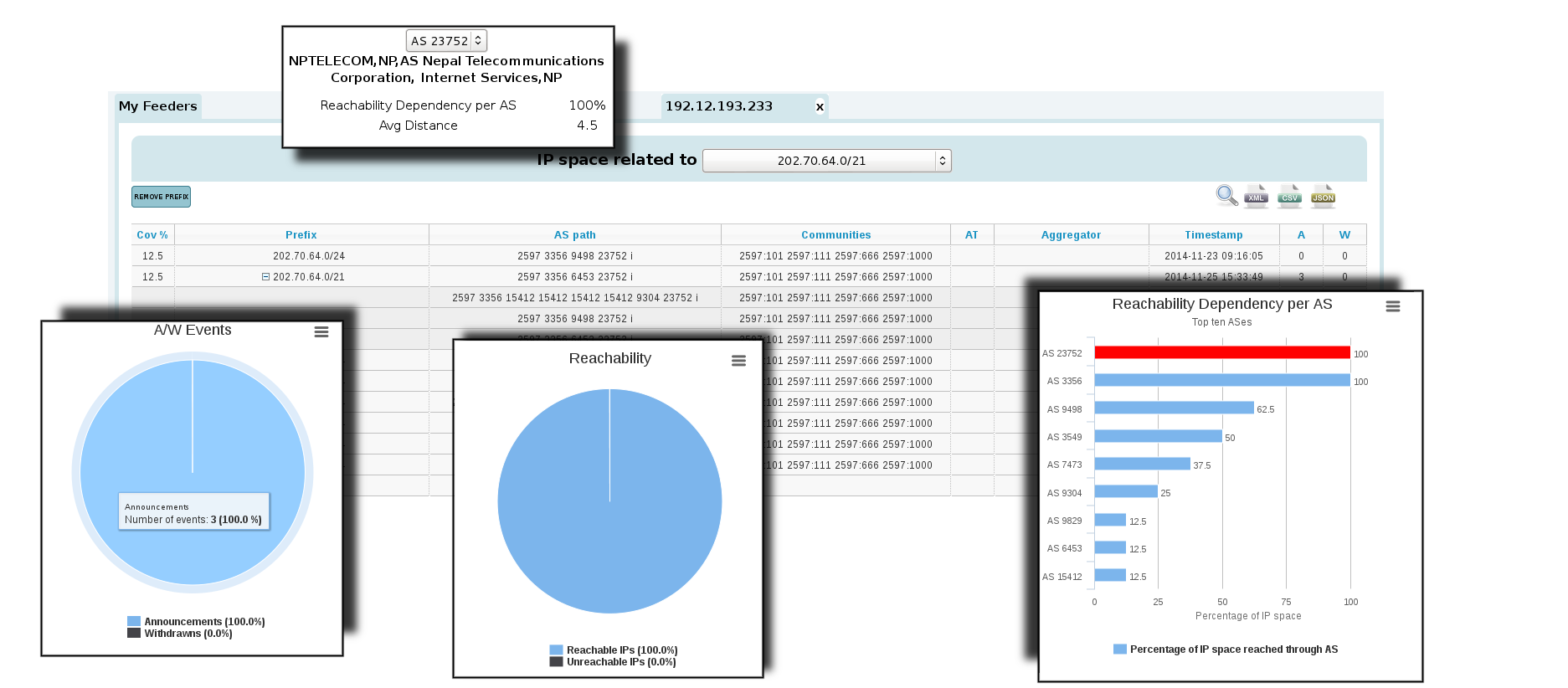}
    \caption{Example of flow-based service: Routing Table Viewer (RTV)}
    \label{fig:rtv}
\end{figure*}

\section{Service overview}
\label{sec:apps}

In the eye of any Isolario user, the most important parts of the system are
the services built on top of the architecture described in
Section~\ref{sec:architecture}. Services are indeed the \textit{do} part of the 
\textit{do-ut-des} principle, and thus the main reason of the participation of
the user to the project and the key to attract new potential users. Up to date
we developed four main categories of services, depending on the data source 
exploited: \textit{flow-based} services, \textit{subnet-based} services, 
\textit{historic} services and \textit{alerting} services. Here in this section 
we provide a brief description of each of these categories, together with some 
examples of services already available to use.

\input{flow_based}

\input{subnet_based}

\input{historic}

\input{alerting}

%% file: flow_based.tex
\subsection{Flow-based services}

\begin{figure}[!b]
    \includegraphics[width=\columnwidth]{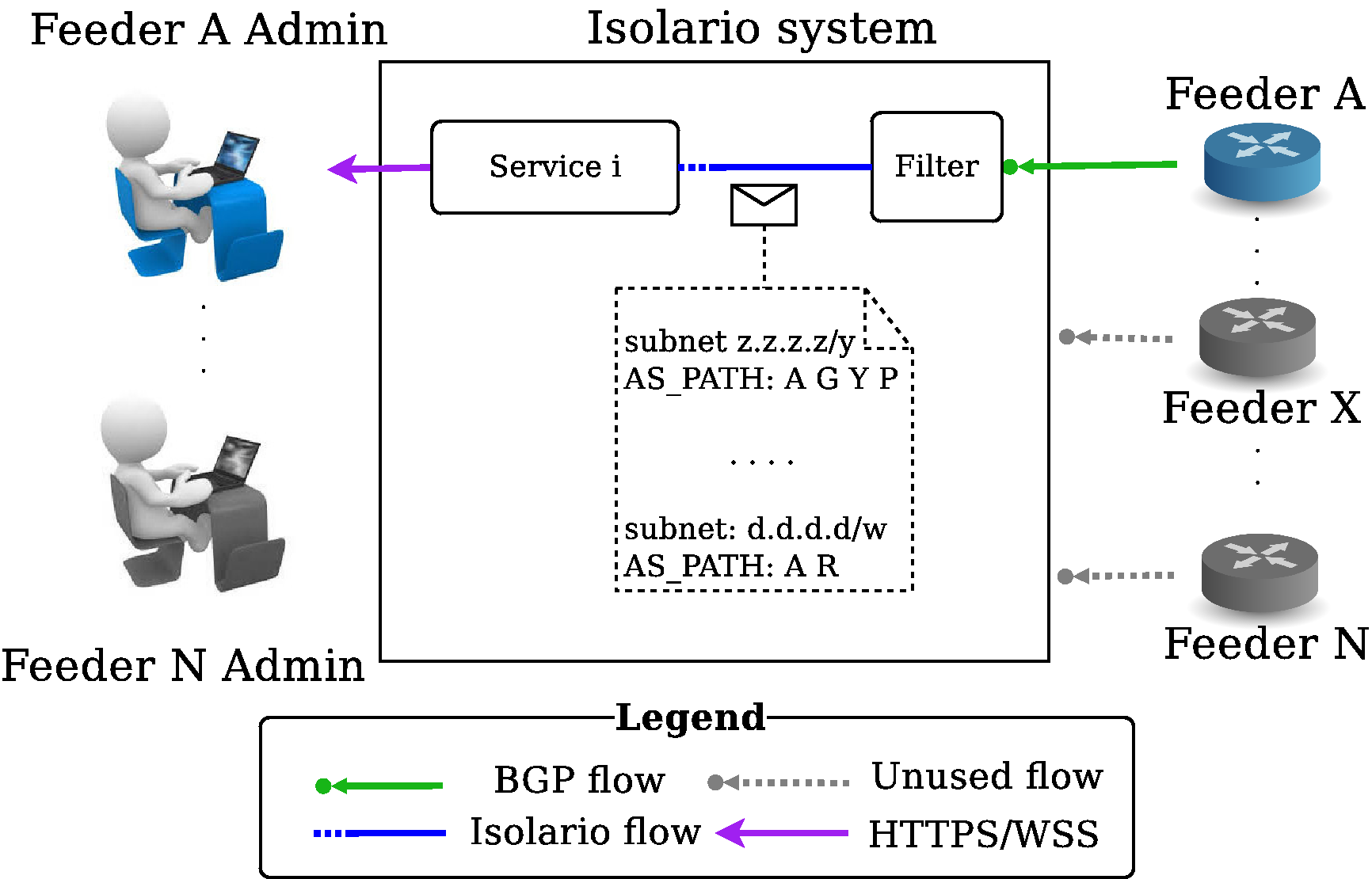}
    \caption{Flow-based services overview}
    \label{fig:flow_based}
\end{figure}

\begin{figure*}[!t]
    \includegraphics[width=\textwidth]{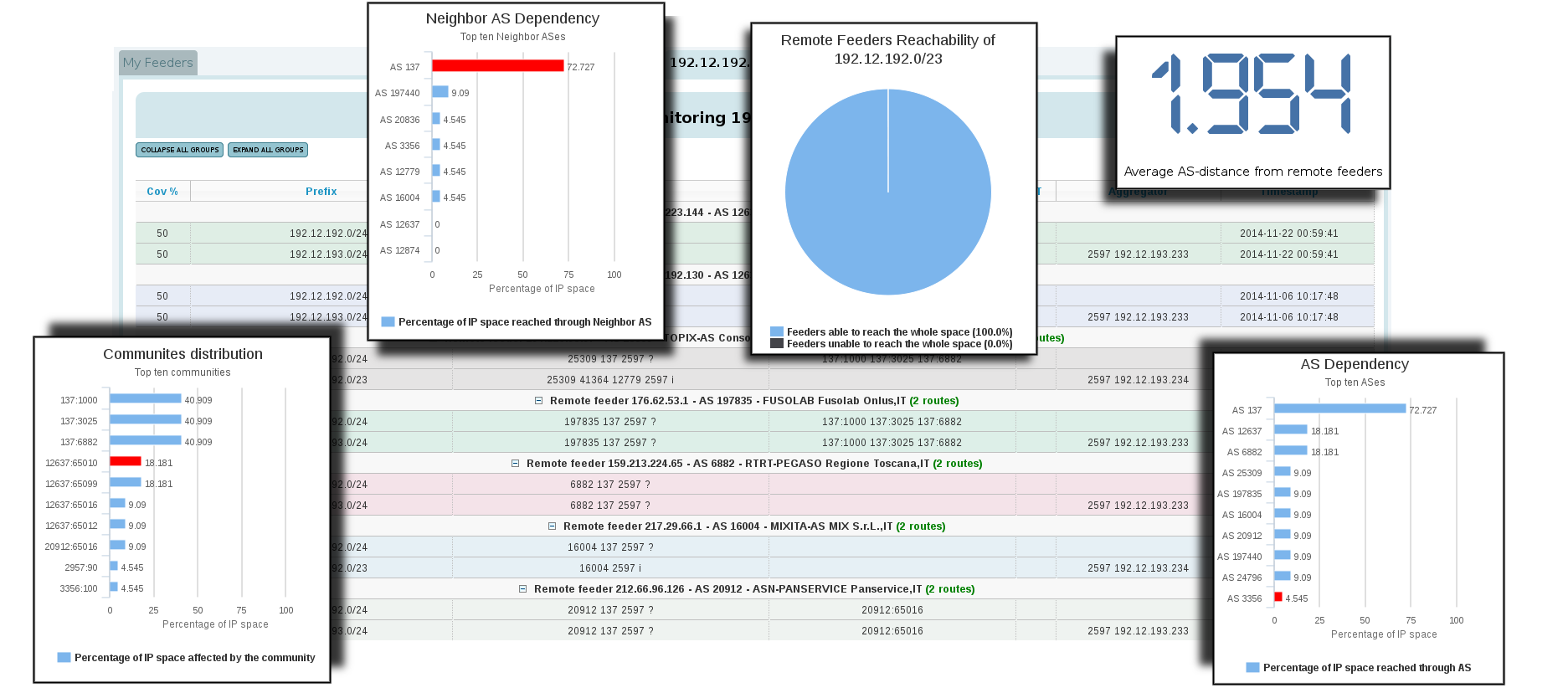}
    \caption{Example of subnet-based service: Subnet Reachability (SR)}
    \label{fig:msr}
\end{figure*}

Some of the possible routing problems that a common network administrator could
face during a normal workday are caused by the behaviour of its BGP neighbours.
For example, network performances could degrade due to high routing 
\texttt{UPDATE} volumes from one of its peers, or some destinations on the
Internet could not be reachable simply because its provider is not announcing 
them anymore (for any reason). In theory a network administrator should monitor the routing 
tables of each ASBR under control 24/7 to understand at glance whether the AS is experiencing these 
problems, but this is not feasible. A possible approach is to perform multiple 
accesses to the ASBRs via Command Line Interface (CLI) trying to understand if 
something is wrong. Another alternative is to use third-party monitoring 
tools -- increasing the CPU load on the ASBR itself -- or to create ad-hoc 
software directly fed by the ASBR -- if resources and know-how are available. 
In practice, the most common way to discover these malfunctions is through the 
complaints of people inside the AS organisation. In the worst case scenario, 
however, the network problem could go unnoticed and could cause a loss of 
revenue, depending on the business of the organisation. Isolario provides a 
valid alternative to that by offering to each Isolario user a set of 
\textit{real-time} monitoring services based on the BGP flow(s) that the user 
organisation established with Isolario, without increasing the CPU load on the 
peering ASBR or requiring the installation of additional software on the 
user-side. \\

The schema followed by this class of services is depicted in 
Figure~\ref{fig:flow_based}. As soon as the user triggers the requests from the
web browser, the related service generates a proper set of filter configuration
messages to allow BGP data of interest to flow in the system. For example, if 
users request to analyse the routing related to \texttt{X.Y.0.0/16}, the 
service will receive in real-time every \texttt{UPDATE} message where
\texttt{X.Y.0.0/16} is explicitly announced or withdrawn, as well as every 
\texttt{UPDATE} message announcing or withdrawing subnets (e.g. 
\texttt{X.Y.1.0/24}) or supernets (e.g. \texttt{X.0.0.0/8}) of the indicated
prefix. The set of filter configuration message and their effect on the system
strictly depend on the needs of each service. These messages are then parsed
and elaborated from the given service module on ERCs and the requested results 
are provided back to the user, together with some related statistics (see for 
example Figure~\ref{fig:rtv}). Note that the parsing and elaboration phase is 
performed in real-time, and that results are shown in real-time on the user web 
browser as well by exploiting the WebSocket protocol. \\

Up to date, every Isolario user can use \textbf{BGP Flow Viewer (BFV)}, which 
shows the \texttt{UPDATE} messages that the Isolario route collector is 
receiving from the ASBR(s) managed by the user, and which allows to quantify the
amount of routing traffic currently generated by the given ASBR(s), \textbf{Routing 
Table Viewer (RTV)}, which allows users to analyse portions of the routing table 
announced by their ASBR(s) -- which coincide with the best routes identified by
the BGP decision process of the ASBR -- and to monitor their evolution, allowing
users to investigate possible reachability problems in real-time (see 
Figure~\ref{fig:rtv}), and \textbf{Route Flap Detector (RFD)}, which allows 
users to detect route flaps\footnote{Route flapping is a rapid sequence of route
state changes which have a heavy impact on router efficiency~\cite{RFC4098}. 
Route flaps are classified in RFD according to the classification of BGP 
dynamics originally defined by Labovitz et al.~\cite{Labovitz98} and then 
redefined by Li et al.~\cite{Li07}.} 
in real-time, allowing them to take the opportune countermeasures.

%% file: subnet_based.tex
\subsection{Subnet-based services}

One of the most (work related) desires of a common network administrator is that
the networks he/she manages are reachable from everyone everywhere at anytime. 
This desire cannot be achieved simply with correct management of the networks 
of the AS managed, and it requires also excellent network planning skills to 
choose the best BGP neighbours and their role (provider or peer). For example, an
AS could benefit in being connected to an Internet Exchange Point (IXP) where
several AS of interest are co-located, and/or in selecting multiple providers
which guarantee different performances in different regions of the world. The 
main problem is that it is extremely hard -- if not impossible -- for any 
network administrator to understand and prove the efficacy of the choice made,
and how these choices impact on the other ASes composing the Internet. For 
example, if AS \texttt{X} peers with AS \texttt{Y} at an IXP, the network
administrator of \texttt{X} cannot assume in any way that \texttt{Y} is going to
use the same IXP to route back all the traffic towards the networks of 
\texttt{X}. This because every single AS has its own perspective of the 
Internet, and there is no collaboration in creating a big picture. The only 
public tools that network administrator could use so far to obtain a 
different perspective about their networks are looking glasses\footnote{A 
\textit{looking glass} is a software run by several ASes to allow public but 
limited access to their BGP routers, in order to ease Internet routing 
troubleshooting.} or MRT data collected by Route Views and RIS. On one side, 
looking glasses are easy to access but can offer just a snapshot of the 
reachability of the network under analysis from the single perspective of the AS
which provide the looking glass access. This means that the network 
administrator would not be able to understand if the networks under analysis are 
experiencing route flaps or if network problems occurred before or after the 
looking glass access. On the other side, MRT data can provide more perspectives
than looking glasses, but their analysis require know-how and often the 
development of dedicated software. Moreover, they can just show the past 
behaviour of network reachability, and not the present. Given its innovative 
structure, Isolario is able to merge the pros of the two approaches \textit{and}
to provide results in \textit{real-time}, thus allowing network administrators 
to check the reachability of their networks with just a couple of clicks. \\

\begin{figure}[!t]
    \includegraphics[width=\columnwidth]{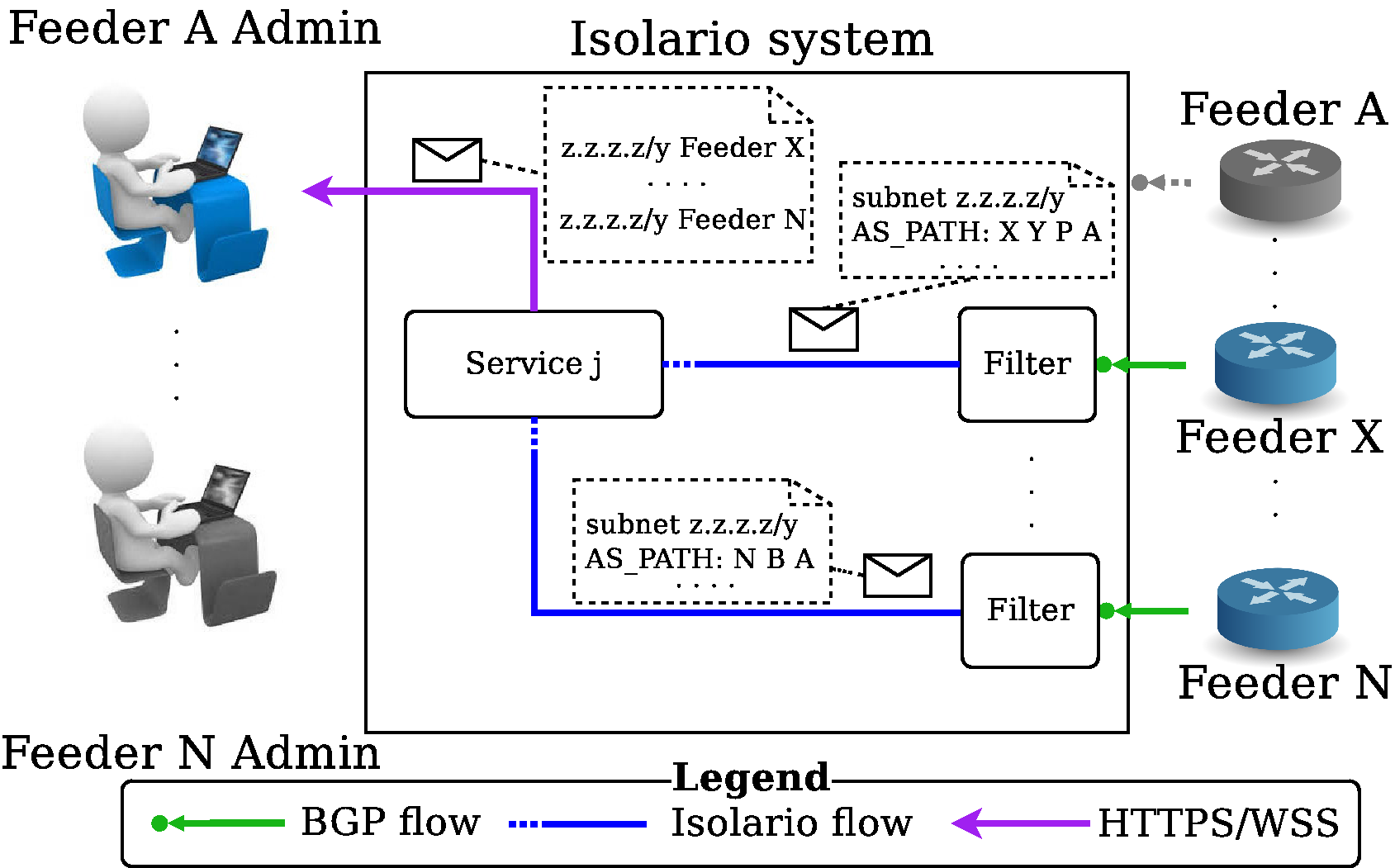}
    \caption{Subnet-based services overview}
    \label{fig:subnet_based}        
\end{figure}

The schema followed by this class of services is depicted in 
Figure~\ref{fig:subnet_based}. Each service generates a set of filter 
configuration messages to allow BGP data to flow in the system but, differently
from flow-based services, these messages are sent to filters regulating 
\textit{every} incoming BGP flow. For example, if the user requires to analyse 
the routing related to \texttt{X.Y.Z.0/24}, the service will receive in 
real-time every \texttt{UPDATE} message where \texttt{X.Y.Z.0/24} is explicitly
announced or withdrawn, as well as every \texttt{UPDATE} message announcing or 
withdrawing subnets or supernets of the indicated prefix from \textit{every} 
incoming BGP flow available in Isolario. These messages are then parsed and 
elaborated from the given service module locally on ERCs and then each partial 
piece of information is merged in the service module in the system core. 
Finally, the requested results are provided to the user, again together with 
some related statistics (see for example Figure~\ref{fig:msr}). So far, the only
service currently available in this class is the self-descriptive \textbf{Subnet 
Reachability (SR)}.

%% file: historic.tex
\subsection{Historic data services}
\label{sec:historic}

Historia magistra vitae\footnote{Latin expression for \textit{History is life's
teacher.}} is an old proverb which means that the study of the past should 
serve as a lesson to the future. This also applies to network administrators,
which may learn from the analysis of \texttt{UPDATE} messages announced in the
past something about the resilience and robustness of their routing system. For 
example, a network administrator would benefit to understand how its 
reachability was dented during a catastrophe like the 9/11 or which network its
routing system was not able to reach during accidental submarine cable cuts or 
malicious cyber attacks. Moreover, a window in the past could help in 
troubleshooting current routing problems and could help in choosing the correct 
peers and providers to diversify the connectivity of their AS. Historic 
information are thus extremely useful, but to date network administrators were
required to analyse MRT data by themselves, or to create their own historic 
repository privately inside their own AS, creating their private route collecting systems.
Isolario exploits HDSRS (see Section~\ref{sec:hdsrs}) to 
provide each user something similar to a time machine. So far, we 
developed the historic aliases of some of the flow-based and subnet-based 
services described above, \textbf{Historic Routing Table Viewer (HRTV)} and 
\textbf{Historic Subnet Reachability (HSR)}. The former service allows users to 
investigate the routing related to a given portion of the Internet in a given 
time period, basing on data they announced to Isolario. The latter allows them 
to investigate the routing related to their own network prefixes during given 
periods of time.

%% file: alerting.tex
\subsection{Alerting services}
\label{sec:alert}

Services described so far assume the physical presence of users, which need to 
actively interact with Isolario via their web browser. Obviously, this is not
always possible. According to Murphy's law, outages and peculiar network events
are more likely to happen when the network administrator is not checking 
real-time services (for whatever reason) than when they are logged in Isolario. 
Needless to say, in several cases it is extremely important that network 
administrators act as fast as possible to fix their network problems to avoid,
for example, huge business losses, security issues and/or a lot of annoyed 
customers. To help them in their work, Isolario devised a special class of 
services aimed at monitoring BGP flows provided by feeders in order to catch and
notify interesting events for users. Such events may be related to BGP packet 
content or to protocol fluctuations -- like route flaps or temporary network 
unreachability -- or may be related to serious security issues, like prefix 
hijack attempts and bogons \cite{Zhang07, Qiu07}. \\

Every service in this class runs 24/7 typically following the rules that each 
user set up in the service web page. Depending on the service chosen, filter 
configuration messages will flow following the schema described in flow-based or
subnet-based services, with the service configuration set by the user. As soon 
as BGP messages of interest reach the system core, the service will identify 
whether an alarm has to be triggered or not. If an alarm is triggered, every 
data related will be retrieved to the interested user both through the Isolario 
website (if connected) and through the communication channels indicated by the 
user in the alarm configuration phase, like for example e-mail and/or HTTP/HTTPS
POST messages. Note that notifications triggered when the interested user is not
online are collected on the web core and made available at its next login. \\

Besides the general alerting features described above, a further interesting example of
service in this class is the \textbf{Prefix Hijack Detector (PHD)}. A prefix is
considered to be hijacked whenever an AS announces, maliciously or not, prefixes
out of its own address space. PHD is currently set up to identify from every
single perspective available in Isolario any attempt of hijack of any subnet
belonging to one of the Isolario feeders and communicate that to the related
user. In particular, PHD is able to detect three main classes of hijack events, 
as described in~\cite{Lad06}: \textit{false origin} hijacks whenever an AS 
announces as its own a network prefix belonging to another AS, \textit{covered} 
and \textit{covering} hijacks whenever an AS announces respectively as its own a
subnet or a supernet of a network prefix belonging to another AS.

%% file: conclusions.tex
\section{Conclusions and future work}
\label{sec:conclusions}

The lack of interest in route collecting project is the major cause of the 
incompleteness of BGP data collected by Route Views and RIS. To increase the 
interest of network administrators in sharing their routing information we 
proposed Isolario, an enhanced route collecting project based on the do-ut-des
principle. To increase the number of BGP data sources, Isolario offers a set 
of useful services to help network administrators in troubleshooting network 
reachability problems in exchange for their IPv4/IPv6 BGP full routing tables. 
These services and the whole Isolario architecture have been developed with
particular care to user experience in order to attract as many network 
administrators as possible. To achieve that, we developed a dedicated
\textit{route collecting engine} which is designed for BGP route collecting
and which allows fast RIB dumps -- thus enabling real-time services to obtain 
the steady routing information of a given subnet as fast as possible -- and a 
\textit{historic data storage and retrieval system} which 
guarantees fast access to routing information to historic services and avoids
repeatedly read operations of the same (possibly large) MRT files. \\
 
Isolario is currently accepting new feeders. MRT data collected from the 
feeders will be made available on the Isolario website (https://www.isolario.it) 
and a free trial of Isolario services -- limited to Isolario AS connectivity (AS 
2598) -- is available to the public. To join Isolario, we require feeders 
to establish (at least) one BGP session with one of our route collectors and to 
announce us their routes towards all the Internet destinations. In the very near
future we plan to increase the number of services available to broaden the 
usefulness of Isolario and to ease the user access to services.